\documentclass[final,1p]{elsarticle}
\makeatletter
\def\ps@pprintTitle{%
 \let\@oddhead\@empty
 \let\@evenhead\@empty
 \def\@oddfoot{\centerline{\thepage}}%
 \let\@evenfoot\@oddfoot}
\makeatother
\usepackage{amsmath}
\usepackage{amssymb}
\usepackage{amsthm}
\usepackage{graphicx}
\usepackage{txfonts} 
\usepackage{subcaption}
\graphicspath{ {figures/} }
\newcommand{\E}{\psi}

\newcommand{\vr}{\mathbf{r}}
\newcommand{\vp}{\mathbf{p}}
\renewcommand{\Re}{\text{Re}}
\renewcommand{\Im}{\text{Im}}
\newcommand{\Ei}{\text{Ei}}
\newcommand{\inp}[2]{\left\langle #1, #2\right\rangle}
\newcommand{\ainp}[2]{\left|\inp{#1}{#2}\right|}
\DeclareMathOperator{\fft2}{\text{fft}_2}
\DeclareMathOperator{\ifft2}{\text{ifft}_2}

\begin{document}
\begin{frontmatter}
\title{A convergent Born series for solving the inhomogeneous Helmholtz equation in arbitrarily large media}
\author{Gerwin Osnabrugge}
\author{Saroch Leedumrongwatthanakun}
\author{Ivo M. Vellekoop\corref{cor1}}
\ead{i.m.vellekoop@utwente.nl}
\cortext[cor1]{Corresponding author}
\address{Biomedical Photonic Imaging Group, MIRA Institute for Biomedical Technology \& Technical Medicine, University of
Twente, PO Box 217, 7500 AE Enschede, The Netherlands}

\begin{abstract}
We present a fast method for numerically solving the inhomogeneous Helmholtz equation. Our iterative method is based on the Born series, which we modified to achieve convergence for scattering media of arbitrary size and scattering strength. Compared to pseudospectral time-domain simulations, our modified Born approach is two orders of magnitude faster and nine orders of magnitude more accurate in benchmark tests in 1-dimensional and 2-dimensional systems.
\end{abstract}

\end{frontmatter}
\section{Introduction}
The Helmholtz equation, the time-independent form of the scalar wave equation, appears in many fields of physics ranging from electron scattering to seismology. Additionally, this equation describes electromagnetic wave propagation in 2-dimensional systems, and it is often used as a scalar approximation to light propagation in 3-dimensional scattering media \cite{rossum1999}.

A great variety of approaches are available for solving the Helmholtz equation in inhomogeneous media \cite{erlangga2008}. For most numerical methods, the equation is discretized using the finite difference \cite{nabavi2007} or the finite element \cite{melenk2010,thompson2006} approximation. The resulting system of linear equations can be solved using direct matrix inversion methods \cite{martinsson2009,schmitz2012}, but these methods are ineffective for solving large inhomogeneous systems. More efficient methods exist, such as the category of Krylov subspace methods \cite{gander2015} extended with multigrid \cite{stolk2014,erlangga2006} or domain decomposition methods \cite{stolk2013,gordon2013}. However, solving the inhomogeneous Helmholtz equation for large systems remains a computationally challenging task \cite{ernst2012}.

For solving Maxwell's equations, the finite-difference time-domain methods (FDTD) \cite{taflove1995,yee1966} and the pseudospectral time domain methods (PSTD) \cite{liu1997,vay2013,tseng2006} are favored the most \cite{gallinet2015}. From these time-domain methods, the steady-state solution is found by replacing the source term with a periodically oscillating source and running the simulations until steady state is reached. However, the finite difference approximations required for these methods are only exact in the limit of an infinitely small step size, resulting in a trade-off between accuracy and speed. 

A completely different class of methods is formed by volume integral methods utilizing the Green's function \cite{beylkin2009,wubs2002,kristensen2010,martin1994}. From Green's function theorem, the well-known Born series can be derived \cite{rossum1999}. The Born series has long been used to solve the scattering problem for small particles with a low scattering potential. However, the series diverges for larger structures, or structures with a high refractive index contrast \cite{kleinman1990}.

Here we present a modified version of the Born series, that unconditionally converges to the exact solution of the Helmholtz equation for arbitrarily large structures with an arbitrarily high scattering potential. Our method converges rapidly and monotonically to the exact solution, and is several orders of magnitude faster than PSTD. 

First, we will introduce our modified Born series, and discuss the convergence criteria and other characteristics. Then, we compare the accuracy and run-time of our method to that of PSTD for 1-D and 2-D media. A proof of convergence is given in Appendix A and the boundary conditions for our experiments are presented in Appendix B.

\section{Algorithm}\label{sec:algorithm}
We aim to solve the inhomogeneous Helmholtz equation
\begin{equation}\label{eq:inhomogeneous-wave-equation}
\nabla^2 \E(\vr) + k(\vr)^2 \E(\vr) = -S(\vr),\\
\end{equation}
where $\E(\vr)$ represents the field at position $\vr \in \mathbb{R}^{D}$, with $D$ representing the dimensionality of the problem. $S(\vr)$ is the source term and $k(\vr)$ is the wavenumber. We proceed to define the scattering potential $V(\vr) \equiv k(\vr)^2 - k_0^2 - i\epsilon$ to find
\begin{equation}\label{eq:inhomogeneous-wave-equation-split}
\nabla^2 \E(\vr) + (k_0^2 + i\epsilon) \E(\vr) = - V(\vr) \E(\vr) -S(\vr).
\end{equation}
Typically, one chooses $k_0$ as the background potential and $\epsilon$ infinitesimally small \cite{beylkin2009}. However, the solution of Eq.~\eqref{eq:inhomogeneous-wave-equation-split} does not depend on the choice for $k_0$ and $\epsilon$ at all. In our approach, both are positive real numbers that we are free to choose conveniently. 

Using the Green's function theorem, the solution can be written as
\begin{equation}\label{eq:green-function-applied-real-space}
\E(\vr) = \int g_0(\vr-\vr') [V(\vr') \E(\vr') + S(\vr')] d \vr',
\end{equation}
where the Green's function $g_0$ is defined as the solution to
\begin{equation}\label{eq:green-function-definition}
\nabla^2 g_0(\vr) + (k_0^2 + i\epsilon) g_0(\vr) = -\delta(\vr).
\end{equation}
The Green's function can easily be found in Fourier transformed coordinates $\vp$, where $\tilde{g_0}(\vp)=1/(|\vp|^2-k_0^2-i\epsilon)$. Eq.~\eqref{eq:green-function-applied-real-space} can now be simplified to the matrix form
\begin{equation}\label{eq:non-preconditioned} 
\E = G V \E + G S,
\end{equation}
where we replaced the convolution with $g_0$ by a single operator $G\equiv F^{-1} \tilde{g_0}(\vp) F$, with $F$ and $F^{-1}$ the forward and inverse Fourier transforms, respectively. $V$ is a diagonal matrix, and $S$ and $\E$ are vectors. Eq.~\eqref{eq:non-preconditioned} can be expanded recursively to arrive at the traditional Born series
\begin{equation}\label{eq:original-Born}
\E_\text{Born} = [1 + G V + G V G V + \ldots] G S.
\end{equation}
Note that this series only converges to $\E$ if the spectral radius $\rho$ of operator $GV$ is less than unity, which is generally not the case. However, here we show that the series \emph{can} be made convergent by choosing a suitable preconditioner $\gamma$. Applying this preconditioner to Eq.~\eqref{eq:non-preconditioned} gives
\begin{align}
\gamma\E &= \gamma G V \E + \gamma G S,\\
\E &= M \E + \gamma G S,\label{eq:preconditioned}
\intertext{with}
M &\equiv \gamma G V - \gamma + 1.
\end{align}
We now choose
\begin{align}
\gamma(\vr)   & = \frac{i}{\epsilon}V(\vr),\label{eq:convergence-condition-gamma}\\
\epsilon & \geq {\max}_{\vr} |k(\vr)^2-k_0^2|.\label{eq:convergence-condition-epsilon}
\end{align}
As we show in \ref{sec:appendix_convergence} this combination of $\gamma$ and $\epsilon$ ensures that $\rho(M)<1$, meaning that Eq.~\eqref{eq:preconditioned} can be expanded recursively to arrive at our modified Born series
\begin{equation}\label{eq:modified-Born}
\E = [1+M+M^2+M^3+\ldots] \gamma G S,
\end{equation}
or, in iterative form
\begin{equation}\label{eq:modified-Born-iteration}
\E_{k+1}=M\E_k + \gamma G S,
\end{equation}
with $\E_0 = \gamma G S$. This iteration converges to the exact solution of the Helmholtz equation for arbitrarily large media.

\subsection{Interpretation}
In order to gain an intuitive understanding of our algorithm, we first analyze the role of $\epsilon$. As can be seen from Eq.~\eqref{eq:green-function-definition}, $\epsilon$ introduces an imaginary component to the wavevector of the background medium. As a result, the Green's function in an homogeneous medium decays exponentially with distance. For instance, in three dimensions the solution to Eq.~\eqref{eq:green-function-definition} is
\begin{equation}\label{eq:Green-3d}
g_0(\vr) = \frac{e^{i |\vr| \sqrt{k_0^2+i\epsilon} }}{4 \pi |\vr|}.
\end{equation}
The exponential decay makes that the total amount of energy represented by the Green's function is finite and localized.

The term $i\epsilon$ in the background potential is compensated exactly by an imaginary term in the scattering potential $V$, such that the final solution of the Helmholtz equation remains the same. A physical interpretation of this construct is that the background medium is lossy, and the scattering potential compensates this loss by an equal amount of gain. As a result, even a homogeneous medium has a non-zero scattering potential.

\begin{figure}
	\centering	
	\begin{subfigure}[b]{0.55\textwidth}
		\includegraphics[width=\textwidth]{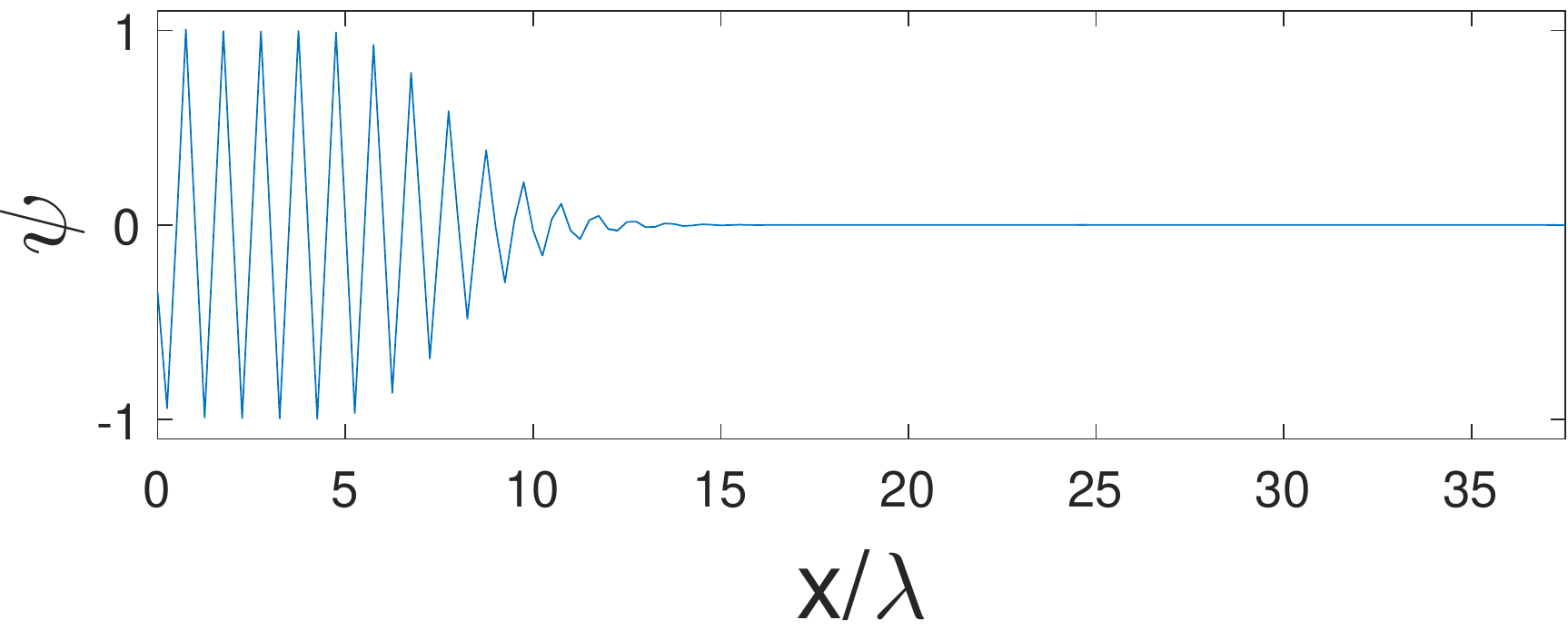}
		\caption{}
	\end{subfigure}	
	
	\begin{subfigure}[b]{0.55\textwidth}
		\includegraphics[width=\textwidth]{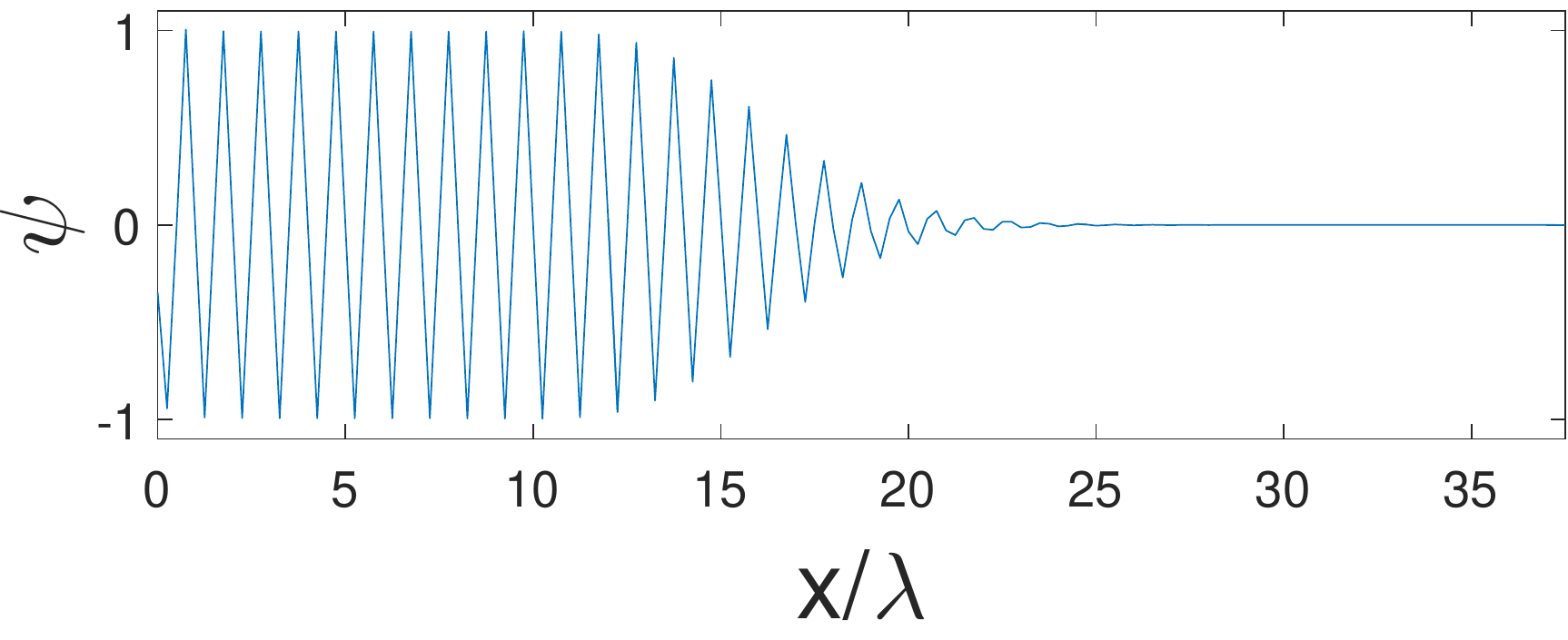}
		\caption{}
	\end{subfigure}	
	
	\begin{subfigure}[b]{0.55\textwidth}
		\includegraphics[width=\textwidth]{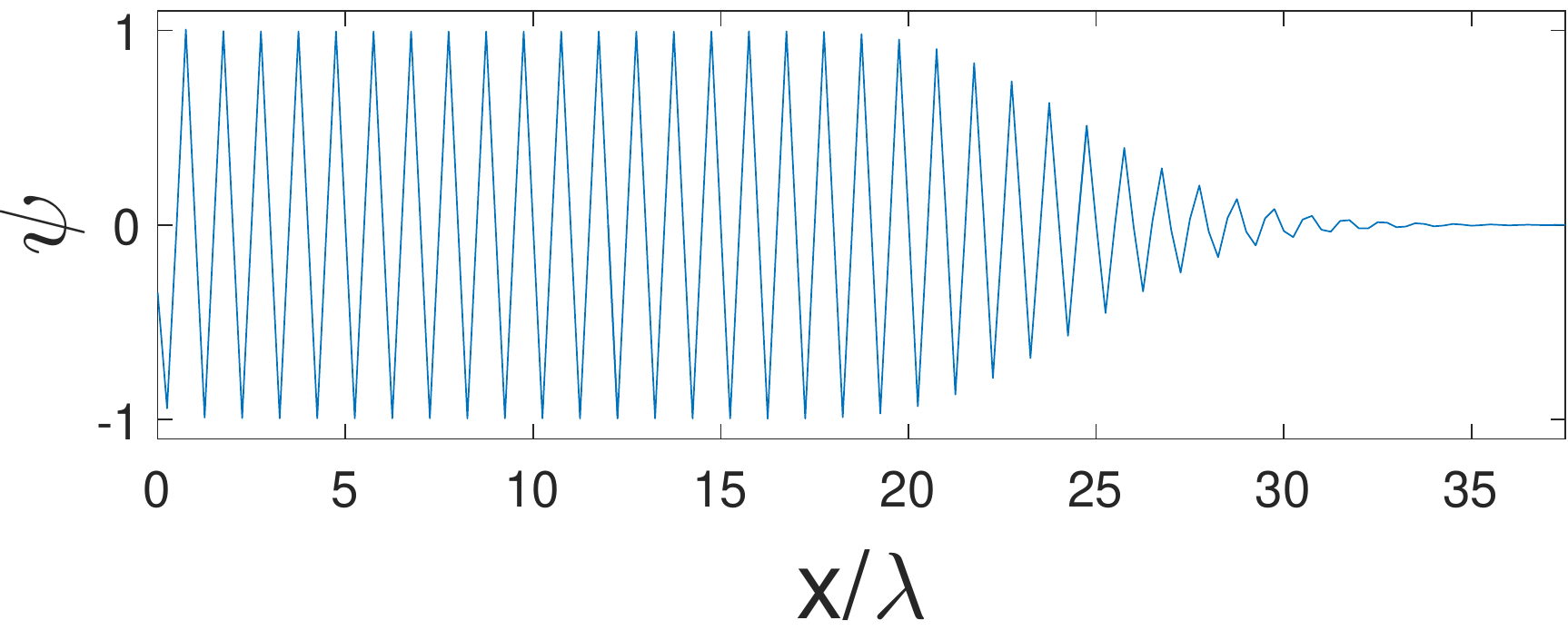}
		\caption{}
	\end{subfigure}
	
	\begin{subfigure}[b]{0.55\textwidth}
		\includegraphics[width=\textwidth]{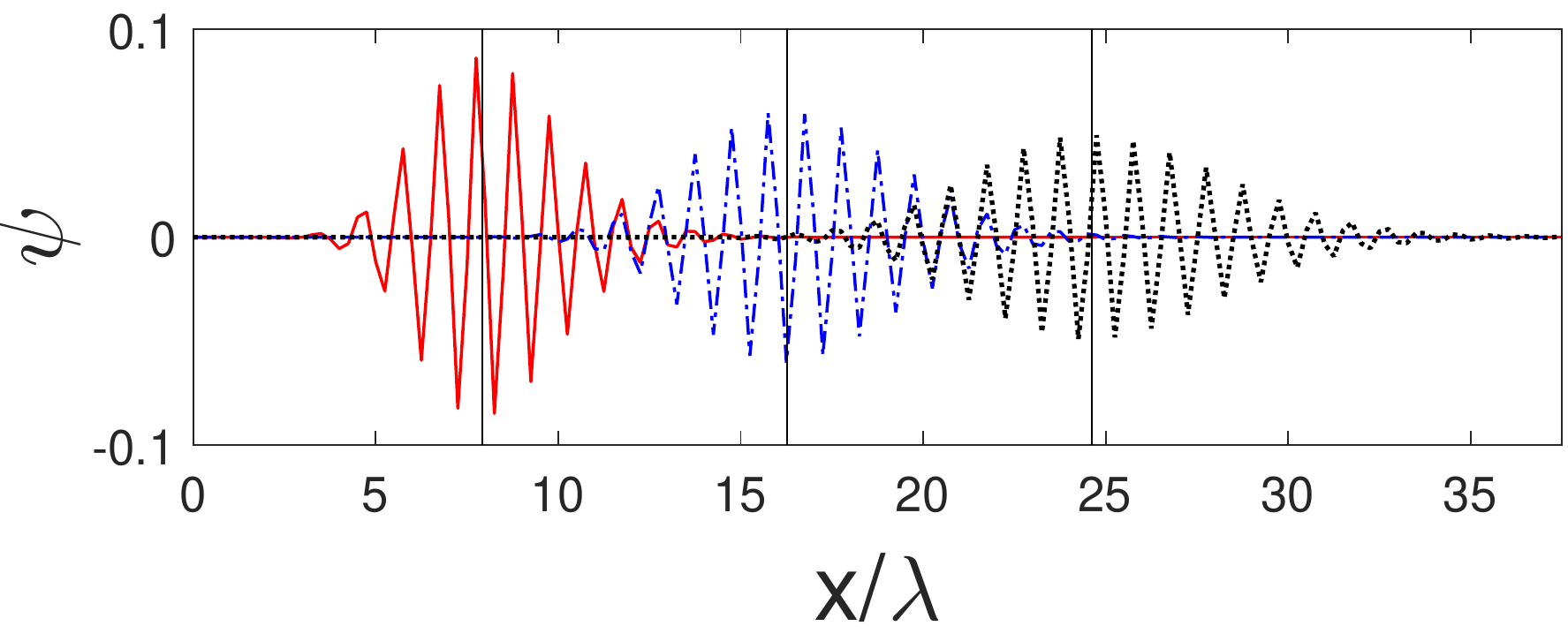}
		\caption{}
	\end{subfigure}	
	
	\caption{a, b, c) solution after 20, 40, and 60 iterations respectively. d) field added during iteration 20 (solid red), iteration 40 (dashed blue), and iteration 60 (dotted black). The vertical lines show the theoretical positions of the differential contributions, calculated using Eq.~\eqref{eq:pseudo-propagation}. In all simulations $\epsilon = 0.8 k_0^2$.}.
	\label{fig:pseudo-propagation}
\end{figure}

Figures~\ref{fig:pseudo-propagation}a-c show the solution for a homogeneous medium after 20, 40, and 60 iterations respectively. It can be seen that the solution expands with each iteration. This expansion should not be confused with time-resolved propagation, since the algorithm is only solving a time-independent wave equation. Still, it is useful to analyze this `pseudo-propagation' effect since it gives an indication of how many iterations are required to cover a homogeneous medium of a given size. We define the pseudo-propagation distance as the distance traveled by a 1-dimensional wave packet $\Psi_0(x)\equiv A(x) \text{exp}(ik_0 x)$, with $A$ as a positive-valued envelope function. The center of the wave packet is defined as
\begin{equation}
\langle x_0 \rangle
\equiv \frac{\int x A(x) dx}{\int A(x) dx}
=      \frac{\int x \Psi_0(x) e^{-ik_0 x} dx}{\int \Psi_0(x) e^{-ik_0 x} dx}
=      \frac{d\tilde{\Psi}_0(p_x)}{\tilde{\Psi}_0(p_x) i d p_x} \Bigg |_{p_x=k_0}
\end{equation}
with $\tilde{\Psi}_0$ the Fourier transform of $\Psi_0$. After a single iteration the packet has transformed into
\begin{align}
	\Psi_1(x) &= \left[\gamma G V - \gamma +1\right] \Psi_0(x).\\
\intertext{For a homogeneous medium ($V=-i\epsilon$, and $\gamma=iV/\epsilon=1$)}
	\tilde{\Psi}_1(p_x) &= \left[ \frac{-i\epsilon}{p_x^2-k_0^2-i\epsilon}\right] \tilde{\Psi}_0(p_x)
\end{align}
The center of the shifted wave packet $\Psi_1$ is at
\begin{equation}\label{eq:pseudo-propagation}
	\langle x_1 \rangle = \frac{d \tilde{\Psi}_1(p_x)}{\tilde{\Psi}_1(p_x) i d p_x} \Bigg |_{p_x=k_0}
	= 2 k_0 / \epsilon + \langle x_0 \rangle,
\end{equation}
with $\langle x_0 \rangle$ the center of the original wave packet. In words, each iteration moves the wave packet over a distance $2 k_0 / \epsilon$. This behavior is illustrated in Fig.~\ref{fig:pseudo-propagation}d. The plot shows the differential contributions of the 20th, 40th, and 60th iteration. Clearly, each iteration only modifies the field in a finite volume, and the solution expands at a speed of $2 k_0 / \epsilon$ per iteration. Eq.~\eqref{eq:pseudo-propagation} shows that $\epsilon$ should be chosen as low as possible to ensure rapid convergence. Since the minimum value of $\epsilon$ is given by Eq.~\eqref{eq:convergence-condition-epsilon}, our method will be faster for structures with a lower contrast. Therefore, we let $k^2_0=\left({\min}_{\vr}\;\Re\{k^2(\vr)\} + {\max}_{\vr}\;\Re\{k^2(\vr)\}\right)/2$ in order to minimize $\epsilon$.
\begin{table}
	\centering
	\renewcommand{\arraystretch}{1.5}
	\caption{Key parameters for our modified Born approach, and PSTD and FDTD simulation methods. Here the smallest wavelength $\lambda_m \equiv 2 \pi / \max_\vr |k(\vr)|$, $\lambda_c \equiv 2 \pi / k_0$ and L is the length of the system.}
	\begin{tabular}{ l | c | c | c | c }
		 & symbol & Modified Born & PSTD \cite{liu1997} & FDTD \cite{yee1966,taflove1995}\\
		\hline
		grid spacing & $\Delta x$ &
		$\leq \lambda_m / 2$ &
		$\leq \lambda_m / 2$ &
		$\leq \lambda_m / 8$\\
		
		time step & $\Delta t$ &
		n/a &
		$\leq \frac{2\Delta x}{\pi c \sqrt{D}}$ & 
		$\leq \frac{\Delta x}{c \sqrt{D}}$\\ 

		(pseudo-)propagation & $\Delta r$ &
		$ 2k_0 / \epsilon \leq \lambda_m \frac{1}{\pi \left(\frac{\lambda_c}{\lambda_m}-\frac{\lambda_m}{\lambda_c}\right)}$ &
		$c \Delta t \leq \lambda_m\frac{1}{\pi\sqrt{D}}$ &
		$c \Delta t \leq \lambda_m\frac{1}{8\sqrt{D}}$ \\
		
		accuracy &  &
		machine precision &
		$O(\Delta t^2)$ &
		$O(\Delta t^2 \Delta x^2)$\\
		
		fft / iteration & &
		2 &
		2 &
		n/a\\
		
		total grid points & $N_\text{tot}$ &
		$(L/\Delta x)^D$ &
		$(L/\Delta x)^D$ &
		$(L/\Delta x)^D$\\
		
		FLOPs / iteration & &
		$O(N_\text{tot} \log_2 N_\text{tot})$ &
		$O(N_\text{tot} \log_2 N_\text{tot})$ &
		$O(N_\text{tot})$\\
	\end{tabular}
	\label{tab:key-properties}
\end{table}
The pseudo-propagation speed can be compared to the actual propagation speed in other methods such as PSTD and FDTD. Table ~\ref{tab:key-properties} summarizes the key characteristics of our method compared to PSTD and FDTD. By comparing the (pseudo-)propagation speed, it can be seen that if
\begin{equation}
\left(\frac{\lambda_c}{\lambda_m}-\frac{\lambda_m}{\lambda_c}\right)<\sqrt{D},
\end{equation}
our method is always faster than PSTD. For $D=3$, this condition corresponds with $\lambda_c / \lambda_m \leq 2.2$ or $\max_\vr |k(\vr)|/k_0 < 2.2$. For example, if the scalar wave approximation is used to simulate light propagation in biological tissue, with a refractive index between 1.33 and 1.45, our exact method `propagates' at a speed of 2.7 wavelengths per iteration, which is a factor $21$ faster than a PSTD simulation ran at the lowest accuracy possible.

\section{Implementation}
We implemented the modified Born algorithm in Matlab. The field $\E$, and potential map $V$ are discretized on a regular 2-dimensional grid with a grid spacing of $\Delta x = \lambda / 4$, with $\lambda$ an arbitrarily chosen wavelength of 1 distance unit. The field $\E$, the potential $V$, and the pre-computed $\tilde{g}_0$ are stored in 2-dimensional arrays. Our iterative algorithm is implemented as
\begin{equation}
\E_{k+1}(\vr) = \E_k(\vr) - \frac{i}{\epsilon}V(\vr) \left(\E_k(\vr)-\ifft2\left[\tilde{g_0}(\vp) \fft2\left[V(\vr) \E_k(\vr) + S(\vr)\right]\right]\right),
\end{equation}
where $\fft2$ and $\ifft2$ are the forward and inverse 2-dimensional fast Fourier transform operators, index $k$ is the iteration number, and $S$ is the source. All multiplications are point-wise multiplications.

In order to compare the accuracy and efficiency of our approach with that of PSTD, we used a PSTD algorithm to solve the time-dependent scalar wave equation for a time-harmonic source \cite{spa2014}
\begin{equation}
\nabla^2 \E(\vr,t) - \frac{1}{c^2(\vr)}\frac{\partial^2 \E(\vr,t)}{\partial t^2} - \frac{\sigma(\vr)}{c^2(\vr)} \frac{\partial \E(\vr,t)}{\partial t} = -S(\vr)e^{-i\omega t},\label{eq:time-harmonic-wave-equation}
\end{equation}
with $\omega$ the angular frequency of the source, $c(\vr)$ the wave velocity, and $\sigma(\vr)$ the damping rate. We choose the time units to have unit wave velocity, such that $\omega=2\pi/\lambda$. By Fourier transforming Eq.~\eqref{eq:time-harmonic-wave-equation} with respect to time, we find back the Helmholtz equation
\begin{equation}
\nabla^2 \E(\vr) + \frac{\omega^2}{c^2(\vr)}\E(\vr) + \frac{i\omega\sigma(\vr)}{c^2(\vr)}\E(\vr) = -S(\vr).\\
\end{equation}
This result shows that we can solve the Helmholtz equation using PSTD by choosing $c^2(\vr)=\omega^2/\Re\{k^2(\vr)\}$ and 
$\sigma(\vr)=\omega\Im\{k^2(\vr)\}/\Re\{k^2(\vr)\}$, and propagating the simulated wave until a steady state is reached.

We use the following PSTD algorithm \cite{spa2014} to solve the time-dependent scalar wave equation in presence of attenuation
\begin{equation}
\E_{k+1}(\vr) = C^{(1)}(\vr) \E_{k-1}(\vr) + C^{(2)}(\vr) \E_k(\vr) + C^{(3)}(\vr) \left(\ifft2\left[-|\vp|^2 \fft2\left[\E_k(\vr)\right]\right] + S(\vr)\right),     
\end{equation}
where coefficient matrices $C^{(1)}$, $C^{(2)}$ and $C^{(3)}$ are pre-calculated:
\begin{equation}
C^{(1)}(\vr) = \frac{\sigma(\vr) \Delta t - 2}{\sigma(\vr) \Delta t + 2},\qquad
C^{(2)}(\vr) = \frac{4}{\sigma(\vr) \Delta t + 2},\qquad
C^{(3)}(\vr) = \frac{2 c^2(\vr)\Delta t^2}{\sigma(\vr) \Delta t + 2}.
\end{equation}
Since PSTD uses a finite-difference scheme to approximate the time derivative, the accuracy of the method will depend on $\Delta t$. In the limit of $\Delta t\rightarrow 0$, PSTD should converge to the exact solution that is found with our modified Born approach.

We have the option to either use absorbing boundaries or periodic boundary conditions. Choosing a proper absorbing boundary condition is not trivial as any residual reflection or transmission will affect the accuracy of the simulation. We found that for our high-accuracy simulations a good trade-off between layer thickness and accuracy was achieved with the boundaries described in ~\ref{sec:boundary-conditions}.

\section{Results}\label{sec:results}
\subsection{Accuracy and efficiency for homogeneous media}
In the first numerical experiment, we simulate wave propagation through a homogeneous medium with $k(\vr)=2\pi/\lambda$. The simulation grid has a size of 1 by 200 pixels, with a resolution of $\Delta x=\lambda/4$.  A source with unit amplitude was placed at the first pixel of the homogeneous medium. We used periodic boundary conditions in the vertical direction, and appended a 25-$\lambda$-thick absorbing layer (of type $N=4$, see Appendix~\ref{sec:boundary-conditions}) on both the left side and the right side of the medium. The resulting structure of 1 by 400 pixels was zero-padded to a 1 by 512 pixel grid to increase the performance of the fast Fourier transforms. We simulated wave propagation through this medium. First using our modified Born approach, and then using PSTD with various values for $\Delta t$.

For each simulation, we calculated the relative error $E$ with respect to the analytic solution
\begin{equation}
E \equiv \frac{\langle |\E - \E_a|^2 \rangle}{\langle |\E_a|^2 \rangle},
\label{eq:relative_error}
\end{equation}
where $\E_a$ is the exact analytic solution and $\langle \cdot \rangle$ denotes averaging over all pixels in the medium, excluding the boundaries. To calculate the $\E_a$, we represent the single-pixel source as a sinc function, and convolve this source with the Green's function for a homogeneous medium. For simplicity, we only evaluate the convolution for $x>0$
\begin{align}
\E_a(x) &= \frac{\sin(\pi x/\Delta x)}{\pi x/\Delta x} * \frac{e^{i k |x|}}{-2 i k}\\
&= 
\frac{\Delta x}{4 k \pi}e^{-i k x} \left[2 i \pi + \Ei(i k^- x) - \Ei(i k^+ x)\right] +
\frac{\Delta x}{4 k \pi}e^{i k x} \left[\Ei(-i k^- x) - \Ei(-i k^+ x)\right],
\label{eq:analytical-solution}
\end{align}
with $k^-\equiv k-\pi/\Delta x$, $k^+\equiv k+\pi/\Delta x$, and $\Ei$ as the exponential integral function. The solution rapidly converges to $\Delta x \exp(i k x)/(-2 i k)$ with increasing distance from the source. Still, with the accuracy that our method achieves, the contribution of the exponential integrals ($0.06\%$ of the total energy in the simulation) cannot be neglected.

\begin{figure}
\centering
\includegraphics[width=0.55\textwidth]{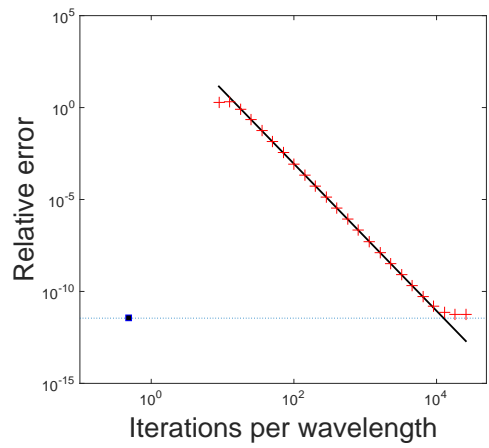}
\caption{ Relative error as a function of a number of iterations per wave cycle for our modified Born approach (square) and PSTD with varying time-step (+). The solid line is proportional to $\Delta t^4$, with $\Delta t$ the PSTD time step. The dotted line corresponds to the accuracy achieved with our modified Born method.}\label{fig:accuracy_homogeneous}
\end{figure}

The result of the simulations are shown in Fig.~\ref{fig:accuracy_homogeneous}. We ran all simulations long enough to traverse 100 wavelengths, i.e. the complete medium twice. The figure shows that our method achieves a relative error $E$ of less than $10^{-11}$ at an exceptionally low computational effort of only $0.5$ iteration per propagated wavelength. For PSTD, we varied $\Delta t$ from the largest possible value allowed by the convergence criterion (see Table~\ref{tab:key-properties}), decreasing the time step until the relative error stabilized. The results of PSTD for large time steps are completely off ($>100\%$ relative error). To achieve an accuracy of 1\% over the full medium, at least 55 iterations per propagated wavelength are needed. The execution speed of both our method and PSTD are limited by the two $\fft2$'s per iteration, so the $x$-axis of Fig.~\ref{fig:accuracy_homogeneous} is directly proportional to simulation time. 

This simple test shows that our method is two orders of magnitude faster, and at least nine orders of magnitude more accurate than a PSTD simulation that is designed to achieve a 1\% accuracy. As can be seen in the figure, the accuracy of PSTD can be increased by increasing the number of steps per wave cycle. Even though the accuracy is proportional to $\Delta t^4$, it takes an extreme value of well over $10^4$ iterations per wave cycle before the accuracy of PSTD matches that of our method. The last data point in Fig.~\ref{fig:accuracy_homogeneous} took 1.5~days to calculate, compared to 0.3~seconds for our modified Born series.

The main source of error in PSTD simulations is an accumulating error in the phase of the propagating wave due to an error in the calculated wavevector \cite{liu1997}. Under some circumstances (e.g. near-paraxial wave propagation) this error may not be a cause of concern. However, in most interesting scenarios (such as resonators, scattering media, integrated optics devices, etc.) the phase needs to be calculated with a high accuracy in order to correctly simulate multi-path wave interference. Therefore, this simulation, where only 50 wave cycles were simulated, may even give an optimistic picture of the accuracy of PSTD.

The relative error of $10^{-11}$ that is found for our method is caused by residual reflection and leakage of the absorbing boundaries, not by the algorithm itself. By choosing boundaries that are thicker and smoother, an error of less than $10^{-17}$ was reached. However, since such extreme accuracies are rarely required, we used thinner boundaries, resulting in a smaller medium and shorter run times.
			
\subsection{Disordered inhomogeneous medium}
In the second numerical experiment, we simulate wave propagation through a 2-dimensional inhomogeneous medium with a random complex potential. The relative scattering potential $k^2(\vr)/k_0^2-1$ is normally distributed with mean $(1.30 + 0.05i)^2$ and variance $(0.10 + 0.02i)^2$. The distribution is low-pass filtered with a cut-off at $1.0\lambda^{-1}$. The simulated medium has a size of 256 by 256 pixels with a resolution of $\Delta x = \text{min} \{ \lambda / 4 \}$ in both spatial dimensions. Periodic boundaries are used and as a source term we take a point source in the center of medium.

\begin{figure}
\centering
\begin{subfigure}[b]{0.49\textwidth}
	\includegraphics[width=\textwidth]{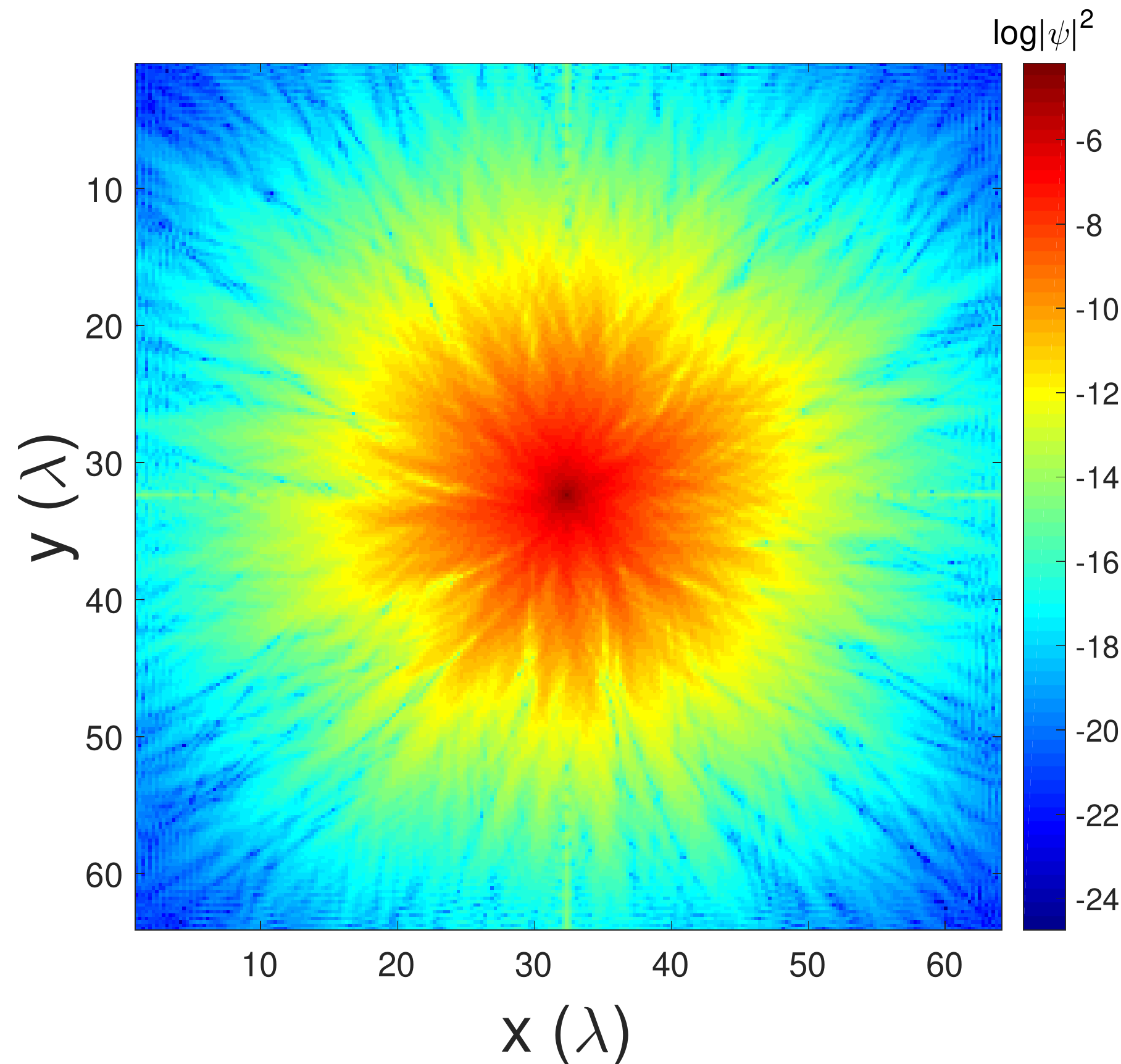}
	\caption{}
	\label{fig:random_solution}
\end{subfigure}	
\begin{subfigure}[b]{0.49\textwidth}
	\includegraphics[width=\textwidth]{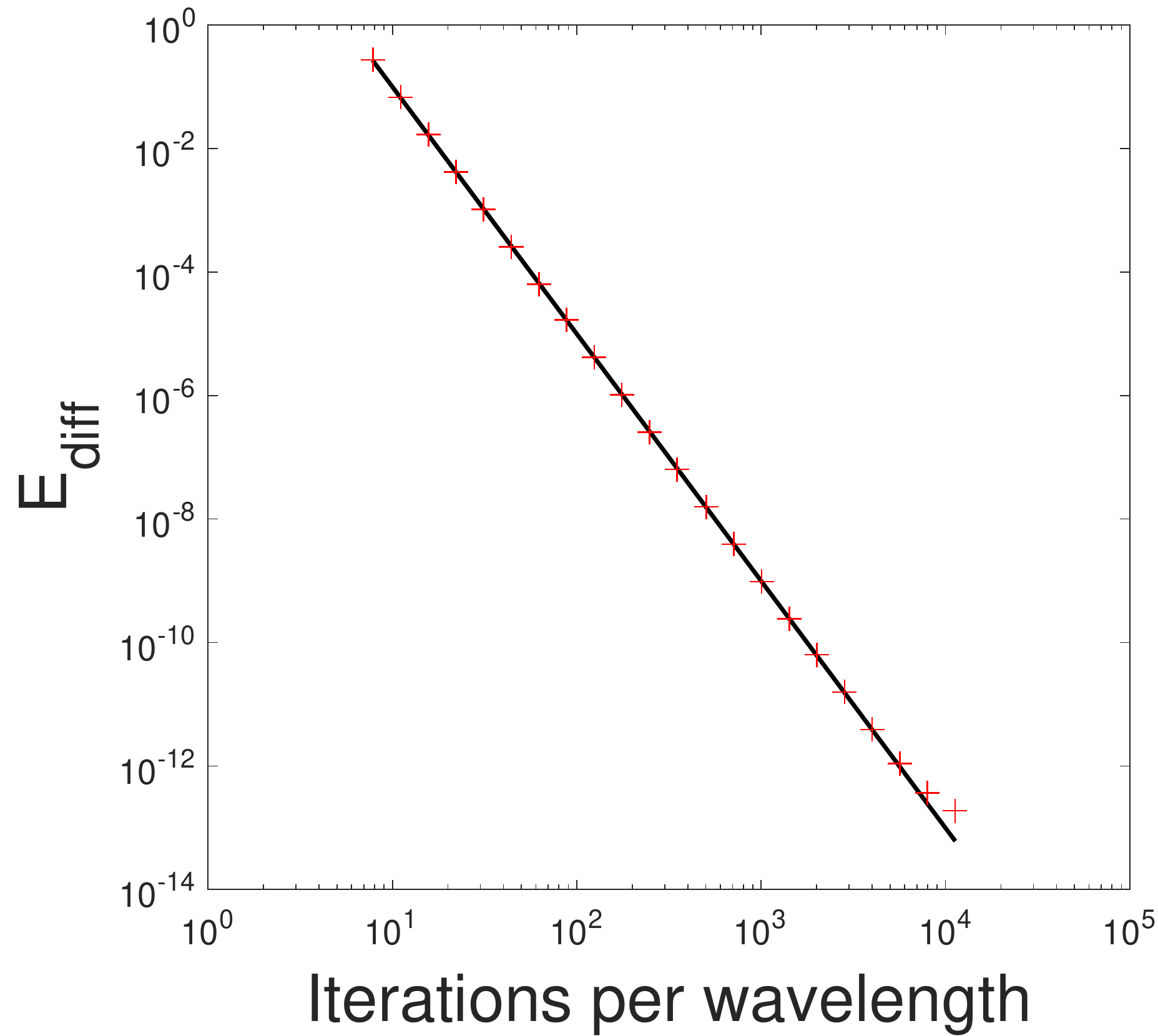}
	\caption{}
	\label{fig:random_results}
\end{subfigure}
	
\caption{Wave propagation of a point source at the center of an inhomogeneous medium with a random potential distribution. a) steady-state solution found using our modified Born approach. b) relative difference between the solution found using the two numerical methods as function of the number of iteration used for the PSTD method. The solid line is proportional to $\Delta t^4$.}
\label{fig:random_medium}
\end{figure}

First, we simulate wave propagation in the inhomogeneous medium using our method. The solution is presented in Fig.~\ref{fig:random_solution}. Then, the numerical experiment was performed using PSTD with various numbers of iterations per wave cycle. Similar to in Eq.~\ref{eq:relative_error}, we define a relative difference:
\begin{equation}
E_\text{diff} \equiv \frac{\langle |\E_\text{PSTD} - \E_\text{Born}|^2 \rangle}{\langle |\E_\text{Born}|^2 \rangle},
\end{equation}
where $\E_\text{PSTD}$ and $\E_\text{Born}$ are the solutions found using PSTD and our method respectively. $E_\text{diff}$ is shown in Fig.~\ref{fig:random_results} as function of the number of iterations per wave cycle used for the PSTD simulations. The results indicate that the solution found by our method is identical to the solution found by PSTD in the limit $\Delta t\rightarrow 0$, to an accuracy of at least $10^{-13}$. Beyond this point, the accuracy of the PSTD simulation started to saturate because of accumulating rounding errors. One should note that the numerical experiment was completed in 8 seconds by the modified Born series method, whereas the PSTD method with $10^4$ iterations per wavelength took 20 hours to finish. This experiment demonstrates that our modified Born series method is highly superior to the PSTD method in terms of both speed and accuracy.

\subsection{Phase conjugation numerical experiments in adipose tissue}
In the last numerical experiment, we simulate a phase conjugation experiment inside a two-dimensional adipose tissue sample of 725 by 725 $\mu$m in size. Previously, such sizes were too time-consuming to simulate, for instance simulating phase conjugation in a 240 $\mu$m thick medium took one week \cite{tseng2009pstd}, whereas our method took 15 minutes to simulate a 725 $\mu$m thick medium.

The refractive index distribution of the adipose tissue model is created using a gray scale microscopy image (Young et al. \cite{young2012}). Based on the optical properties presented in the paper by Jacques \cite{jacques2013}, the refractive index of fatty cells and the extracellular fluids are estimated at 1.44 and 1.36 respectively. Given this range of refractive indices, the gray values in the microscopy image were linearly converted to a refractive index map. We choose the wavelength in vacuum to be $\lambda_0 = 1.0\;\mu$m and scale the microscope image such that the pixel size $\Delta x=\lambda_0/4$. The resulting medium size is 2900 by 2900 pixels, to which we added a 25-$\lambda$-thick absorbing boundary layer and zero padded the medium to a size of 4096 by 4096 pixels.

\begin{figure}
\centering
\begin{subfigure}[b]{0.49\textwidth}
	\includegraphics[width=\textwidth]{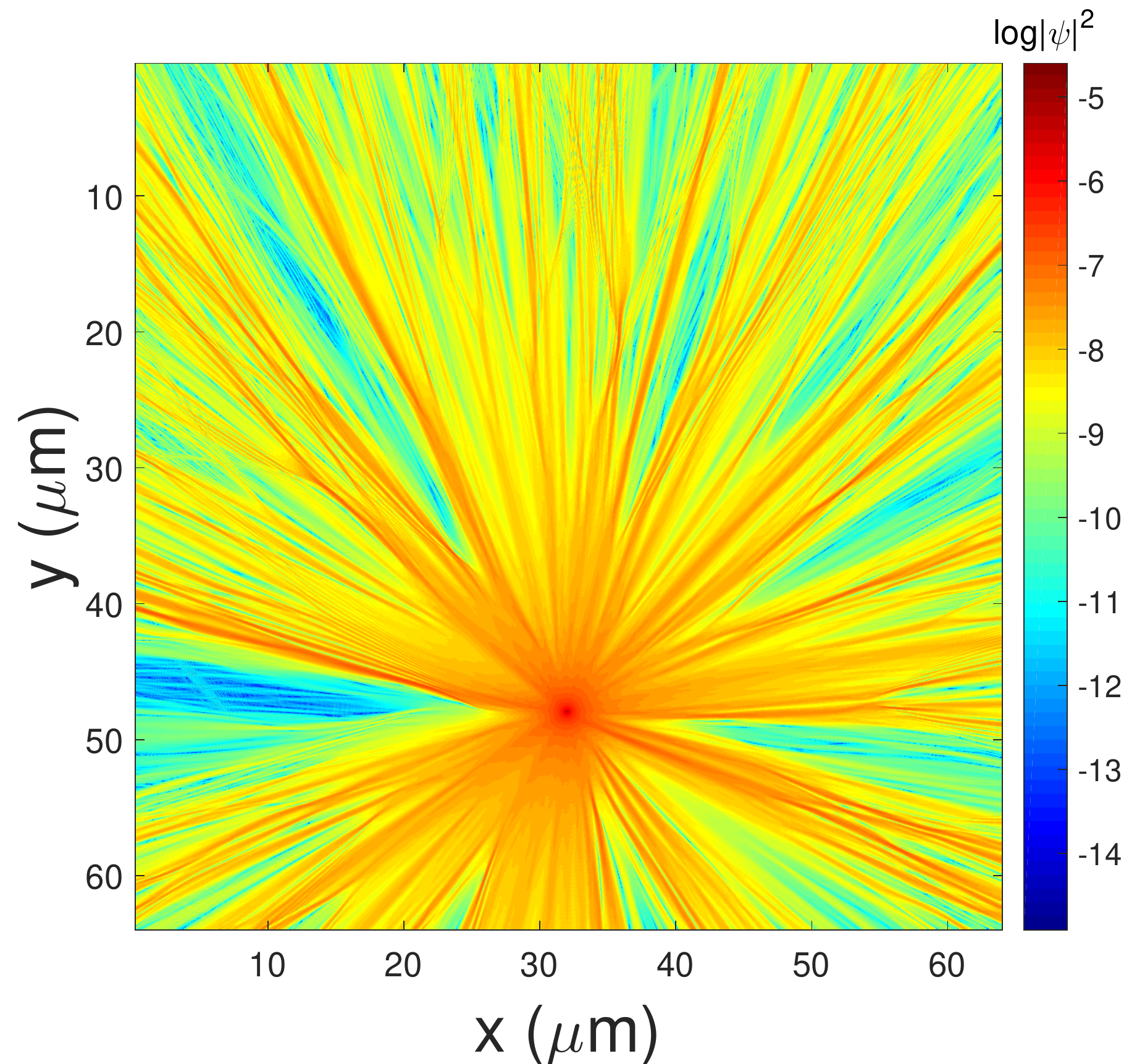}
	\caption{}
	\label{fig:adipose_PS}
\end{subfigure}
\begin{subfigure}[b]{0.49\textwidth}
	\includegraphics[width=\textwidth]{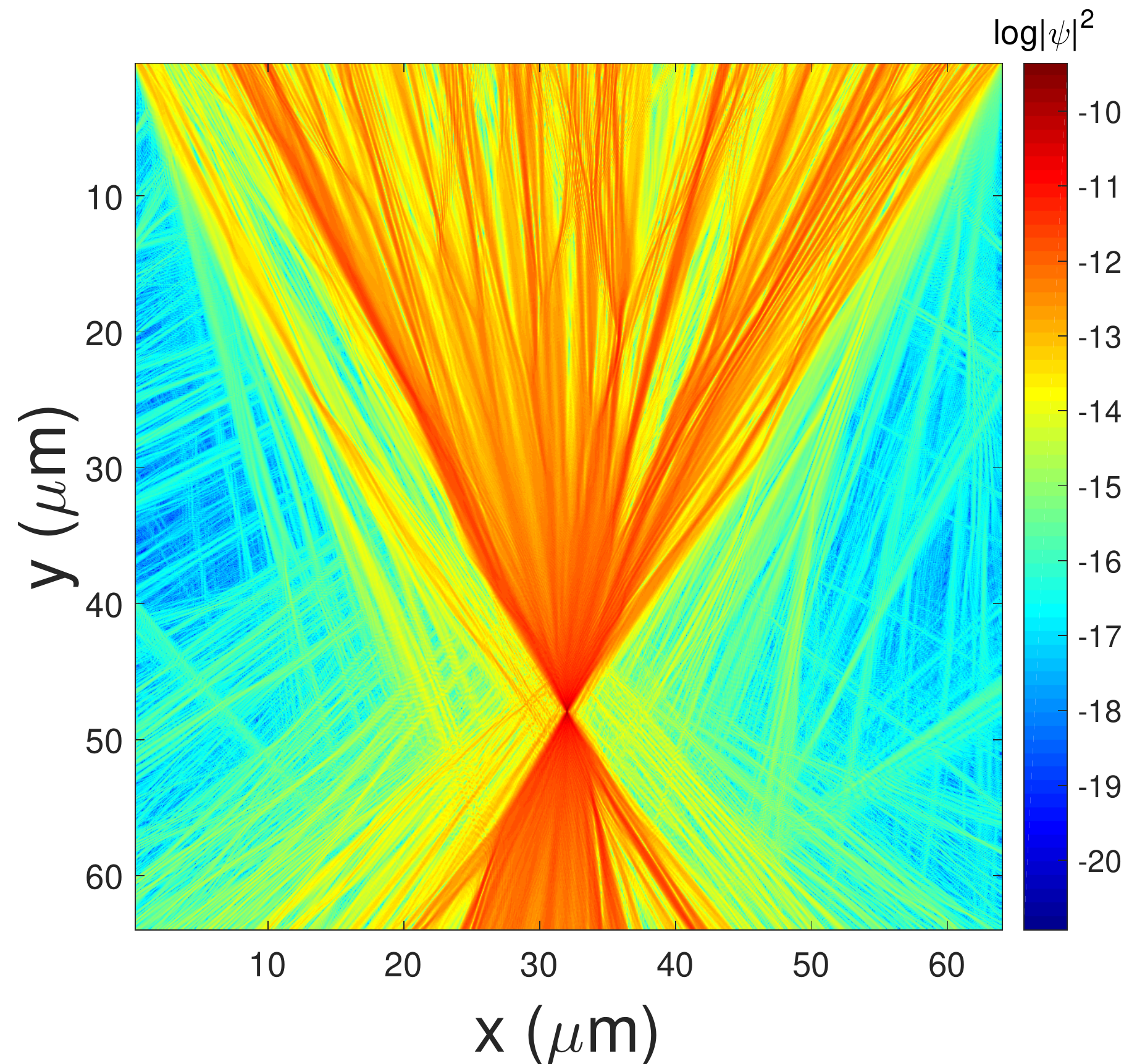}
	\caption{}
	\label{fig:adipose_dopc}
\end{subfigure}
	
\caption{Light propagation inside a two-dimensional adipose tissue model. a) the refractive index distribution based on a gray scale microscopy image of adipose tissue \cite{young2012}. b) steady state solution of a point source inside the tissue. c) phase conjugation of the field at the top boundary, which results in the light focusing at the location of the original point source.}
\end{figure}

We start by placing a point source in the center of the medium, which yields the steady-state solution as shown in Fig.~\ref{fig:adipose_PS}. Afterwards, we take the phase conjugate of the resulting field at the top boundary of the simulation grid and use it as the source of a second simulation. In Fig.~\ref{fig:adipose_dopc}, the steady-state solution of this second simulation is shown. As can be seen in the figure, the light is still scattered as it propagates deeper inside the tissue, however, now it focuses at the location of the point source of the first simulation. 

\section{Conclusion}
In conclusion, we presented a fast iterative method for solving the Helmholtz equation. Our method is a modified version of the Born series. In contrast to the original Born series, our method converges for arbitrarily large structures with an arbitrarily high scattering potential. 

Our method is several orders of magnitude faster in finding a steady-state solution to the inhomogeneous wave equation than the commonly used PSTD and FDTD methods, especially for media with a low scattering contrast. In addition, our method converges to the exact solution, whereas finite difference methods are, by definition, limited in accuracy. 

If a time-dependent solution is required, our modified Born series method could be applied once for each frequency of interest, and the time-dependent solution could be calculated simply with a Fourier transform. Such an approach would still be more accurate, and potentially faster than PSTD.

We demonstrated our method for 1-dimensional and 2-dimensional media. Extensions to scalar waves in higher dimensions are trivial. For electromagnetic wave calculations, the method will need to be modified for vector waves. For such an approach, we expect to find similar improvements in accuracy and speed as for scalar waves.

Future research may be directed towards finding combination of $\gamma$ and $\epsilon$ for which the iterations converge more rapidly, or towards using more advanced iteration methods to speed up convergence. Additionally, due to the localized nature of our Green's function, a large medium may be split into separate domains to accommodate parallel processing or multi-domain calculations.

\appendix

\section{Proof of convergence}\label{sec:appendix_convergence}
In this appendix, we will show that the series in Eq.~\eqref{eq:modified-Born} always converges for the choice of $\gamma$ and $\epsilon$ as given by Eqs.~\eqref{eq:convergence-condition-gamma} and \eqref{eq:convergence-condition-epsilon}. A sufficient condition for convergence is $\rho(M)<1$. We will first prove that $\rho(M)\leq 1$ and consider the eigenvalues with value 1 later. Recognizing that
\begin{equation}
\frac{1}{|\vp|^2-k_0^2-i\epsilon} = \frac{i}{2\epsilon}\left(1-\frac{|\vp|^2-k_0^2+i\epsilon}{|\vp|^2-k_0^2-i\epsilon}\right),
\end{equation}
and substituting Eq.~\eqref{eq:convergence-condition-gamma} we can write operator $M$ as
\begin{align}
M &=  \frac{-V}{2\epsilon^2} \left[1 - F^{-1}\frac{|\vp|^2-k_0^2+i\epsilon}{|\vp|^2-k_0^2-i\epsilon}F\right] V - \frac{iV}{\epsilon}+1,\\
&= \frac{1}{2\epsilon^2}\left[-V^2 + VUV - 2i\epsilon V + 2\epsilon^2\right],
\end{align}
where
\begin{equation}
U \equiv  F^{-1}\frac{|\vp|^2-k_0^2+i\epsilon}{|\vp|^2-k_0^2-i\epsilon}F
\end{equation}
is a unitary operator. In order to prove that $\rho(M)\leq 1$, it suffices to show that $\ainp{x}{Mx} \leq \inp{x}{x}$, for all $x$, where $\inp{\cdot}{\cdot}$ is the inner product. We now use the Cauchy-Schwartz inequality to write $\ainp{x}{VUVx} = \ainp{V^\dag x}{UVx} \leq \sqrt{\inp{UVx}{UVx}}\sqrt{\inp{V^\dag x}{V^\dag x}} = \inp{Vx}{Vx}$, where we used the fact that $V$ is diagonal (and, hence $V^\dag V=V V^\dag$) in the final step. With this result, we can eliminate $U$, resulting in
\begin{equation}
\ainp{x}{Mx} \leq \frac{1}{2\epsilon^2}\ainp{x}{\left[2\epsilon^2 -2i\epsilon V -V^2\right]x}+\frac{1}{2\epsilon^2}\inp{Vx}{Vx},
\end{equation}
To complete the proof, we now need to demonstrate that the right hand side of this equation is never larger than 1. Since $V=V(\vr)$, we require
\begin{equation}\label{eq:convergence-criterion-vr}
\left|2\epsilon^2 -2i\epsilon V(\vr) -V^2(\vr)\right| + \left|V(\vr)\right|^2\leq 2\epsilon^2,
\end{equation}
for all $\vr$. To show that this condition is always fulfilled, we define $\Delta\equiv V+i\epsilon=k^2(\vr)-k_0^2$ and rewrite Eq.~\eqref{eq:convergence-criterion-vr} as
\begin{equation}
\left|\epsilon^2 - \Delta(\vr)^2\right| + |\Delta(\vr)-i\epsilon|^2 \leq 2\epsilon^2,
\end{equation}
which can be written as
\begin{equation}\label{eq:convergence-condition-intermediate}
\left|\epsilon^2 - |\Delta(\vr)|^2 - 2i\Delta(\vr)\Im\Delta(\vr)\right| + |\Delta(\vr)|^2 +\epsilon^2-2\epsilon \Im\Delta(\vr)  \leq 2\epsilon^2.
\end{equation}
A slightly stricter criterion follows from triangle inequality
\begin{equation}\label{eq:convergence-condition-final}
\left|\epsilon^2 - |\Delta(\vr)|^2\right| + 2|\Delta(\vr)|\Im\Delta(\vr) + |\Delta(\vr)|^2 +\epsilon^2-2\epsilon \Im\Delta(\vr)  \leq 2\epsilon^2,
\end{equation}
where we require that $\Im\Delta(\vr)\geq 0$, which means that the medium cannot have any gain. Since we have $\epsilon\geq|\Delta(\vr)|$ from Eq.~\eqref{eq:convergence-condition-epsilon}, condition \eqref{eq:convergence-condition-final} is always fulfilled. Therefore, $\rho(M)\leq 1$.

Eigenvalues of $1$ are only possible for an infinite non-absorbing medium. In this case, the solutions to the Helmholtz equation carry infinite total energy, which means that the solution cannot be found using our method. However, if there are absorbing boundaries, or even if there is a single a finite-size volume with non-zero absorption, there will be some points where the wave is absorbed. In this case the left hand side of Eq.~\eqref{eq:convergence-condition-final} is strictly less than the right hand side, and convergence of our method is guaranteed.

\section{Boundary conditions}\label{sec:boundary-conditions}
In our implementation, we use a fast Fourier transform to evaluate the convolution with the Green's function. As a result, the system has periodic boundary conditions by default. In order to prevent waves to `wrap around' the boundaries, we implement absorbing boundary layers. A wide choice of boundary conditions is available for simulating scalar wave propagation in finite size systems (e. g. \cite{berenger1994,liu1998}). We choose to use a type of absorbing boundaries with the following properties: firstly, the layer is designed to have zero reflectivity for normal incidence. Secondly, the scattering potential $|k^2(\vr)-k_0^2|$ of the layer is bounded to a specified maximum value.

We design the absorbing layers by requiring that the wave has the following form
\begin{equation}
\E(x) \propto P_N(x) e^{i k_0 x -\alpha x},
\end{equation}
with $P_N$ an N'th order polynomial inside the absorbing layer ($x>0$) and equal to 1 for $x\leq 0$. Substituting this desired solution into Eq.~\eqref{eq:inhomogeneous-wave-equation} gives
\begin{equation}
P''_N(x) + 2(i k_0-\alpha) P'_N(x)+\left[k^2(x)-k_0^2-2i k_0\alpha + \alpha^2\right]P_N(x) = 0.
\end{equation}
The simplest solution is found for $N=1$:
\begin{equation}
P_1(x>0) = 1+\alpha x,
\end{equation}
\begin{equation}
k^2(x)-k_0^2 = \frac{\alpha^2 (1 - \alpha x + 2i k_0 x)}{1 + \alpha x}.
\end{equation}
Theoretically, this absorbing boundary has zero reflectivity. However, due to the discretization of the field, waves with a high spatial frequency cannot be represented. These frequency components are required to truthfully represent the waves at the transition from the medium to the absorbing boundary. When these components are missing, some residual reflectivity results. We found that this reflectivity can be reduced by smoothing the boundary. This can be achieved by imposing the constraint that the function $k^2(x)-k_0^2$ and its first $N-2$ derivatives vanish at $x=0$.
A general solution for a polynomial of degree $N$ is given by
\begin{equation}\label{eq:boundary-solution-N}
P_N(x>0) = \sum_{n=0}^N \frac{(\alpha x)^n}{n!}
\end{equation}
\begin{equation}
k^2(x)-k_0^2 = \frac{\alpha^2(N - \alpha x + 2i k_0 x) (\alpha x)^{N-1}}{P_N(x) N!}.
\end{equation}
Choosing a higher value for $N$ results in a smoother boundary with a lower residual reflection. However, it also takes longer for the boundary to reach its maximum absorption coefficient of $\alpha$. In Eq.~\eqref{eq:boundary-solution-N} it can be seen that $P_N(x)$ equals the first $N$ terms of the Taylor expansion of $\exp(\alpha x)$, thereby partially canceling the decay term $\exp(-\alpha x)$ close to the boundary.

The boundary potential saturates at a value of $k^2(x)-k_0^2=-\alpha^2+2ik_0\alpha$ for $x\rightarrow\infty$, irrespective of $N$. Higher values of $\alpha$ result in stronger absorption, at the cost of increasing the scattering potential. Because of condition Eq.~\eqref{eq:convergence-condition-epsilon}, a higher scattering potential results in a higher value of $\epsilon$, which gives a lower pseudo-propagation speed (and hence, slower convergence of the method). Throughout this manuscript used layers with a thickness of 25 wavelengths, with $N=4$ and $\max|k^2(\vr)-k_0^2|=0.2$. These parameters gave a good balance between scattering potential, residual reflectivity, and residual transmission for our high-accuracy simulations.

\bibliographystyle{elsarticle-num}
\section*{References}
\bibliography{ref}

\end{document}